 \documentstyle[12pt]{article}
\begin{document}
 \baselineskip=24pt
 \setcounter{page}{0}
 \baselineskip=25pt \parskip=0pt plus2pt
 \textheight=22cm

 \begin{titlepage}

 \begin{center}
 {\Large\bf
 Fragility and Boson Peak formation in a Supercooled Liquid}
 \end{center}

 \vspace{0.2in}

 \begin{center}
 {\it Sudha Srivastava and Shankar P. Das
 \\ School of Physical Sciences, Jawaharlal Nehru University
 \\New Delhi 110067, India.}

 \end{center}

 \vspace*{1.0in}

 \begin{center}
 {\bf ABSTRACT}
 \end{center}
 \noindent
 We analyze results for the Boson Peak from  the
  neutron time of flight spectroscopy data on Ge-As-Se,
 and Raman spectra  data on m-TCP and OTP,
  using a recent mode coupling model that takes into account the
  coupling of density fluctuations with vibrational modes
  in presence of defects in the supercooled state.
  From the experimental results for different materials
  we observe that for more fragile systems characterized
  by increasing fragility parameter m, a slower relaxation
  of the defect-density correlation is needed to give rise to
  the observed peak in the spectra.

 \noindent

 \vspace{1in}
 \noindent
 PACS number(s) : 64.70P, 05.60, 64.60C
 \end{titlepage}
 \newpage

 %\section{Introduction}
 \noindent
 The extra intensity observed in undercooled liquids
 in the neutron scattering \cite{richter,kanaya}
  as well as in Raman scattering \cite{fischer,sokolov} at low
 frequencies,
 distinct from the quasi-elastic peak is usually referred to as
 the Boson peak in the literature. This characteristics
  feature of the supercooled liquid has been ascribed to
 the coupling between the relaxational and vibrational
 motions in the supercooled liquid in a recent work \cite{pre}
 with self consistent mode-coupling  \cite{kimmaj-adv,sid-trans}
 model.
 The mode coupling theory has a better agreement
  with dynamics in fragile glassy systems.
 The classification \cite{angell} used for glassy systems as
 strong and
  fragile depends on the ease with which structural degradation
  occur in those systems.  A convenient measure for the fragility
  of the system \cite{bohmer} is computation of the slope $m$ of
  the relaxation time $\tau$ curve  against temperature $T$ scaled
  with respect to the corresponding glass transition temperature
 $T_g$,
 \begin{equation}
 \label{m-defn}
 m={{{\rm d log} <\tau >} \over {{\rm d (T_g /T})}}{ | }_{T=T_g}
 \end{equation}
 \noindent
 In Ref. \cite{pre} we have described an extension of the simple
  mode coupling formalism to include the distinct vibrational
  modes that develop at low temperatures in the amorphous state
 for understanding the extra intensity appearing for the
 structure factor.
 The model follows from the
 equations of Nonlinear Fluctuating Hydrodynamics \cite{pre}
  extended to include the defect density in an amorphous
  solid like system which also sustains transverse sound modes.
 Here we have applied the model \cite{pre} to extensive
  data comparison to  understand the underlying  relationship between
the fragility of the system  and the Boson Peak formation.
 We fit the model with respect to the data of
  Russina et. al. from neutron time-of-flight-spectroscopy
 for Ge0.033 As0.033 Se0.934 glass,
 as well as the data  for m-tricresyl phosphate (m-TCP)
 by Sokolov et. al. from Raman scattering.
 The present analysis demonstrates that the
 criteria for the appearance of the peak is crucially related to
 the
 dynamics of defect densities in the disordered system.

 In studying the feedback effects on dynamics due to  slowly
 decaying density fluctuations at supercooled states,
 the memory function $H[\phi(t)]$
 is obtained as a functional of the hydrodynamic correlation
 functions,
 in the following q independent form,
 \begin{equation}
 \label{memory}
 H (t) = c_1 F[\phi_L(t), \phi_T(t)] \psi(t) + c_2 \psi^2 (t)
 \end{equation}
 \noindent
 where $c_1$ and $c_2$ are dimensionless constants determined
 in terms of the wave vector integrals due to the mode coupling
 vertex functions.
 $\phi_L$ and $\phi_T$ are the correlation functions for
 the longitudinal and transverse sound modes.
 The function $F(t)$ is expressed as \cite{pre},
 \begin{equation}
 \label{ansatz}
 F[\phi(t)] = e^{-\delta t} + f(\sigma) \phi_T(t)
 \end{equation}
 \noindent
 where $\delta$ represents the time scale of very slowly
 decaying
 defect density and
 $f(\sigma)=(12-14\sigma)/9(1-2\sigma)$ with
 $\sigma = (3\lambda-2\mu)/[2(3\lambda+\mu)]$  is the Poisson's
 ratio.
 For more details we refer to Ref. \cite{pre}.
 Following the procedure described there
 we have used this model to fit the data of Russina et. al.
 \cite{feri,rita} for Ge-As-Se.
 The central quasi-elastic peak is fitted a Lorentizian of width
 $\delta$,
 and is equivalent to taking the time scales
  of relaxation of the defects to be  same \cite{D+M,dsch} as
 that of the
 final time scale of relaxation of the density fluctuations.
 We show in Fig. 1a the dynamic structure factor as
  a function of frequency.
 The time scale for the decay of the defect density denoted by
 $\delta$  plays a central role in appearance of the peak on the
shoulder
 of the quasi-elastic peak.
  This intermediate peak disappear
  in the shoulder of the quasi elastic peak
  as $\delta$ become large.
 This is shown by the fit of the data of Russina et. al. for a
 higher
 temperature in Figure 1b.
 We analyze the data for temperatures,
 T = 252, 334, 359, 402, 440 and 502 degree Kelvin,
 with suitable values of $c_1$ and $c_2$ in the expression
 (\ref{memory}) for the memory function.
 The parameter $\delta$ which  gives the time scale of relaxation
of defects is used here as the only adjustable parameter.
It is considered in inverse units of the time $\tau_0$ in terms
 of which the MCT equations are expressed in the
 dimension less form. $\tau_o$ can be obtained in terms of the
microscopic frequencies of the liquid state.
To study the behavior of $\delta$ with
the fragility parameter we have considered data on another material
m-TCP with a different $m$, used by Sokolov et. al.
 \cite{sokolov} and done a similar fitting to the Boson peak
 at temperatures T = 205, 235, 262 and 287 degree Kelvin. 
In Fig. 2 the fit to the Boson Peak data \cite{sokolov} for m-TCP is shown.
Variation of log($\delta$) with the inverse
 temperature $T_g/T$ is shown in Fig. 3, where $T_g$ is the glass transition 
temperature of the corresponding material.
 The data points for m-TCP are shown by stars ($\ast$) and for OTP \cite{ss-spd} 
by filled circles ($\bullet$).
 Variation of $log(\delta)$ with temperature
 for Ge-As-Se alloy  is shown in the inset by  squares ($\Box$). 
Solid lines show the straight line fit to the data points.
 For lower temperatures the quantity $\delta$ is small indicating
 that the defect densities are long lived and the Boson Peak
 appears to be more pronounced.
 The more fragile the system is, sharper is the fall of $\delta$
 which
  represents the time scale of defect correlation.
 In all three cases $\delta$ shows an Arrhenius fall with
 temperature 
  and a corresponding activation energy can be computed from
  the slope of the curve.
In figure 4 we show the  plot of activation energy $A$ (in
units of temperature T)
  for different systems with the corresponding fragility index m.
  With the systems of increasing fragility, more long lived
  defects are needed to give rise to the observed peak
  in the spectra.

 We have approximated here through $\delta$
  relaxation of the defect density by single exponential
 mode. The full wave vector dependence has to be considered
 to account the coupling of the structural relaxation to the
 vibrational modes.
The explicit temperature dependence of the peak
      is not captured in the present model.
 This can be computed through proper input for
 the static or thermodynamic properties that appear
 in the mode coupling integrals.
 Figure 3, demonstrates the key result of this paper that for
 more fragile
 systems $\delta$, inverse of which relates to the defect density
 correlation,
 shows a sharper fall with temperature.
 Also we like to point out here that the temperature range
 covered for
 the materials in this paper actually correspond to the part
 where the
 fragile glasses starts showing a sharp increase of viscosity on
 the
 Angell plot\cite{angell}
 - more fragile the liquid is, more dramatic is the increase
 giving
  a higher value for the fragility index m \cite{bohmer,angell1}.

 \vspace*{2cm}

 \section*{Acknowledgements}
 F. Mezei and M. Russsina are acknowledged for providing
 Ge0.033 As0.033 Se0.934 glass,  their data for analyzed in the
 paper.
 A. Sokolov is acknowledged for the data for OTP and m-TCP used
  in the analysis.
SS aknowledges SRF support from UGC.

 \vspace*{1cm}
 \newpage
 \section*{Figure Captions}

 \subsection*{Figure 1}
 \noindent
 The neutron scattering data ( in arb. units) of Ref.
 \cite{feri,rita}
 normalized with respect
 to the Bose factor $\omega[n(\omega)+1]$ ( open circles) at
 (a) $T =252^o$ K and (b) $T =440^o$ K,
 Vs. the frequency in Mev. The solid line presents the result
 obtained from the present model for the normalized correlation
function
 $\psi$.

 \subsection*{Figure 2}
 \noindent
 The Raman Spectra data ( in arb. units)  for TCP
 normalized with respect
 to the Bose factor $\omega[n(\omega)+1]$ ( open circles) at
 $T=205^o$ K,
 vs. the frequency in GHz. The solid line presents the result
 obtained from
 the present model for the normalized correlation function $\psi$.

 \subsection*{Figure 3}
 \noindent
$\rm{log}(\delta)$ as a
 function of temperature $(T_g/T)$ for OTP ($\bullet$) and 
for m-TCP($\ast$). Inset: For Ge-As-Se alloy ($\Box$).
The solid lines represent the straight line fit to the data points.

 \subsection*{Figure 4}
 \noindent
 Slope for the $\delta$-1/T curve, $A$ ( in unit of $^o K$ ) Vs the
fragility
 parameter \cite{bohmer} m.

 
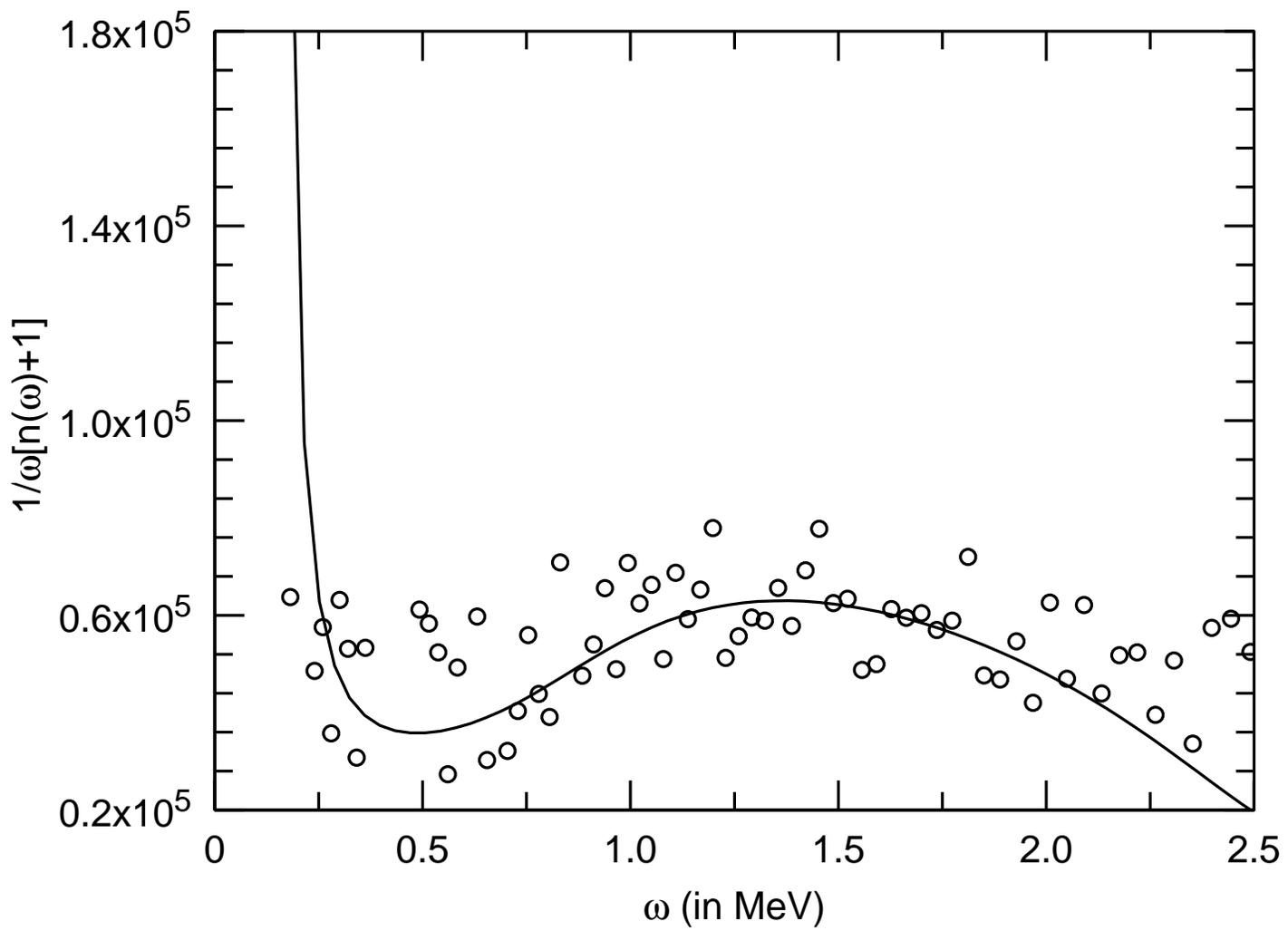
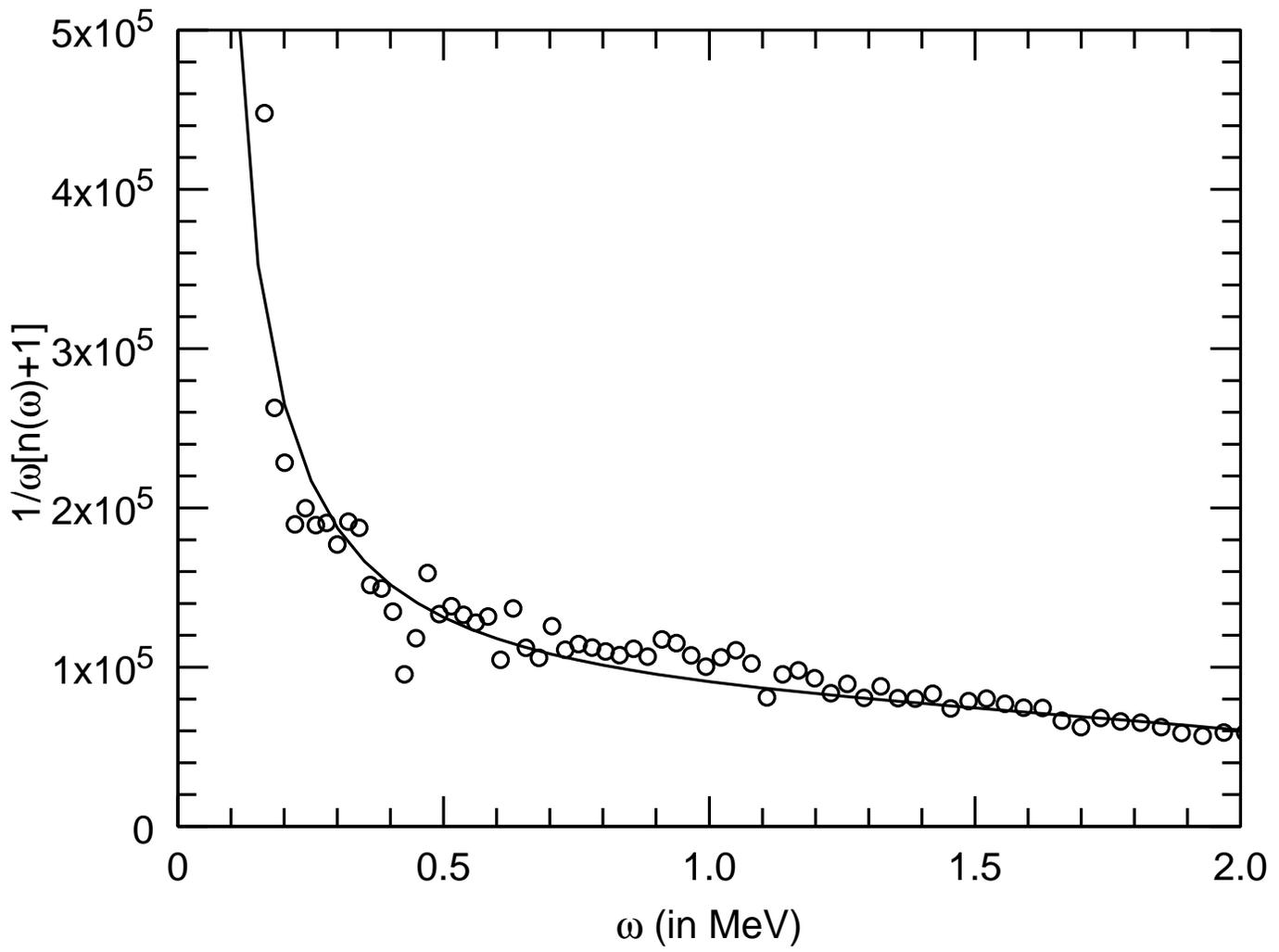
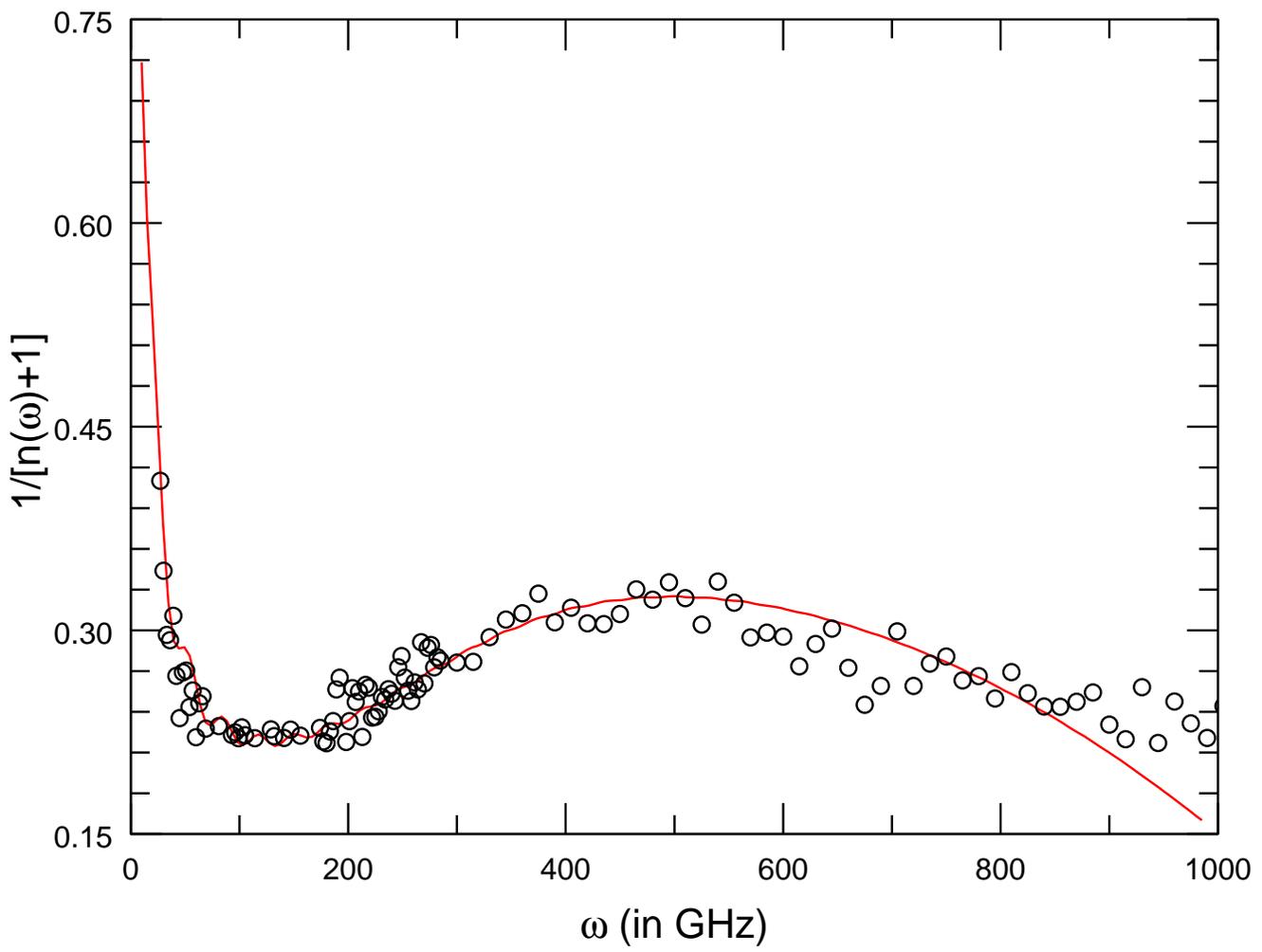
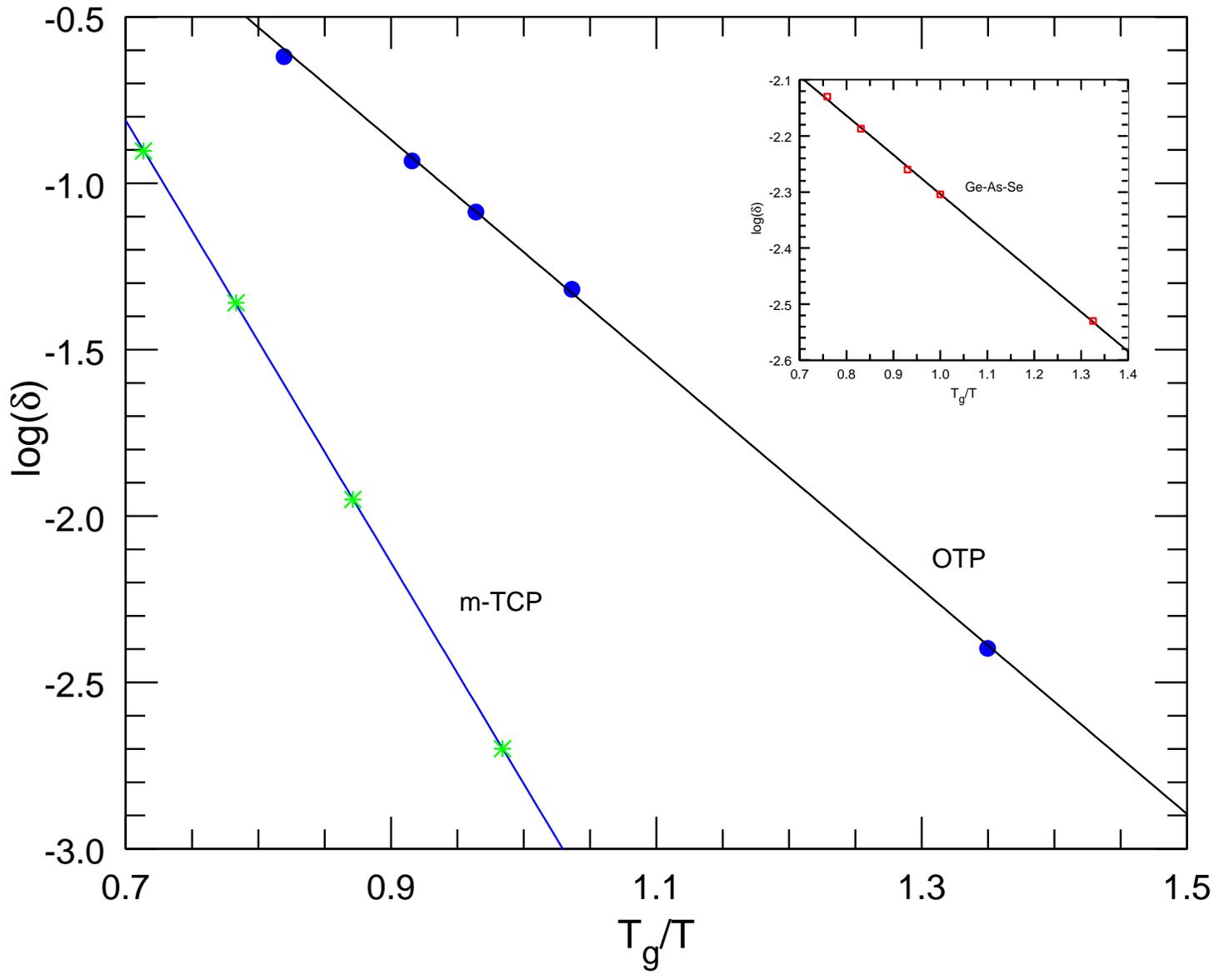
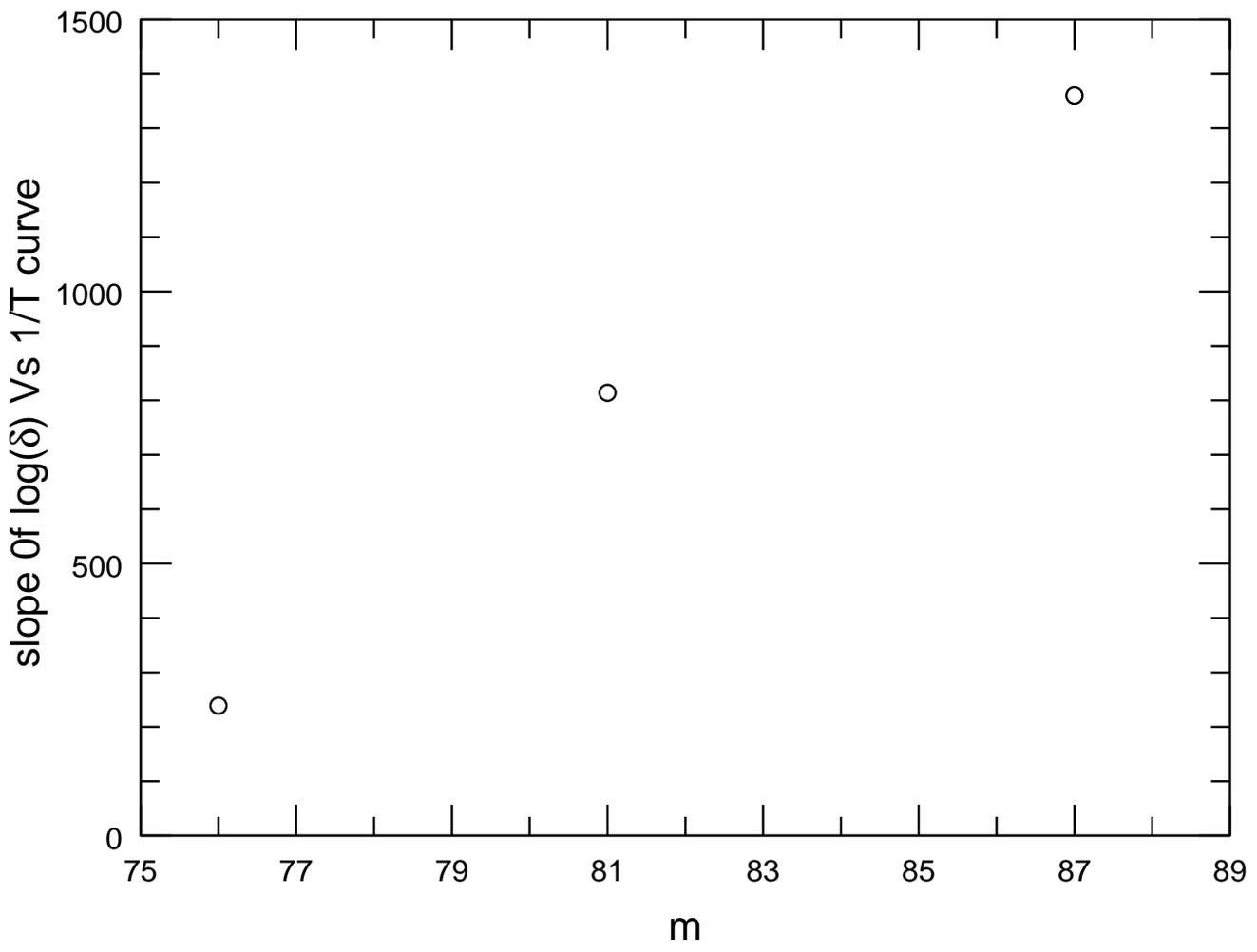
\end{document}